\documentclass[a4paper,twocolumn,english,showpacs,aps,prl,fleqn,superscriptaddress]{revtex4-1}

\usepackage[T1]{fontenc}
\usepackage[utf8]{luainputenc}
\usepackage{amsmath}
\usepackage{amssymb}
\usepackage{graphicx}
\usepackage{hyperref}
\usepackage{babel}
\usepackage{wasysym}
\usepackage{microtype}
\usepackage{color}

\begin{document}
\renewcommand\figurename{FIG.}
\title{Coupled Dipole Oscillations of a Mass-Imbalanced Bose and Fermi Superfluid Mixture}
\author{Yu-Ping Wu}
\affiliation{Shanghai Branch, National Laboratory for Physical Sciences at Microscale and Department of Modern Physics, University of Science and Technology of China, Shanghai, 201315, China}
\affiliation{CAS Center for Excellence and Synergetic Innovation Center in Quantum Information and Quantum Physics, University of Science and Technology of China, Hefei, Anhui 230026, China}
\affiliation{CAS-Alibaba Quantum Computing Laboratory, Shanghai, 201315, China}

\author{Xing-Can Yao}
\affiliation{Shanghai Branch, National Laboratory for Physical Sciences at Microscale and Department of Modern Physics, University of Science and Technology of China, Shanghai, 201315, China}
\affiliation{CAS Center for Excellence and Synergetic Innovation Center in Quantum Information and Quantum Physics, University of Science and Technology of China, Hefei, Anhui 230026, China}
\affiliation{CAS-Alibaba Quantum Computing Laboratory, Shanghai, 201315, China}

\author{Xiang-Pei Liu}
\affiliation{Shanghai Branch, National Laboratory for Physical Sciences at Microscale and Department of Modern Physics, University of Science and Technology of China, Shanghai, 201315, China}
\affiliation{CAS Center for Excellence and Synergetic Innovation Center in Quantum Information and Quantum Physics, University of Science and Technology of China, Hefei, Anhui 230026, China}
\affiliation{CAS-Alibaba Quantum Computing Laboratory, Shanghai, 201315, China}

\author{Xiao-Qiong Wang}
\affiliation{Shanghai Branch, National Laboratory for Physical Sciences at Microscale and Department of Modern Physics, University of Science and Technology of China, Shanghai, 201315, China}
\affiliation{CAS Center for Excellence and Synergetic Innovation Center in Quantum Information and Quantum Physics, University of Science and Technology of China, Hefei, Anhui 230026, China}
\affiliation{CAS-Alibaba Quantum Computing Laboratory, Shanghai, 201315, China}

\author{Yu-Xuan Wang}
\affiliation{Shanghai Branch, National Laboratory for Physical Sciences at Microscale and Department of Modern Physics, University of Science and Technology of China, Shanghai, 201315, China}
\affiliation{CAS Center for Excellence and Synergetic Innovation Center in Quantum Information and Quantum Physics, University of Science and Technology of China, Hefei, Anhui 230026, China}
\affiliation{CAS-Alibaba Quantum Computing Laboratory, Shanghai, 201315, China}

\author{Hao-Ze Chen}
\affiliation{Shanghai Branch, National Laboratory for Physical Sciences at Microscale and Department of Modern Physics, University of Science and Technology of China, Shanghai, 201315, China}
\affiliation{CAS Center for Excellence and Synergetic Innovation Center in Quantum Information and Quantum Physics, University of Science and Technology of China, Hefei, Anhui 230026, China}
\affiliation{CAS-Alibaba Quantum Computing Laboratory, Shanghai, 201315, China}

\author{Youjin Deng}
\affiliation{Shanghai Branch, National Laboratory for Physical Sciences at Microscale and Department of Modern Physics, University of Science and Technology of China, Shanghai, 201315, China}
\affiliation{CAS Center for Excellence and Synergetic Innovation Center in Quantum Information and Quantum Physics, University of Science and Technology of China, Hefei, Anhui 230026, China}
\affiliation{CAS-Alibaba Quantum Computing Laboratory, Shanghai, 201315, China}

\author{Yu-Ao Chen}
\affiliation{Shanghai Branch, National Laboratory for Physical Sciences at Microscale and Department of Modern Physics, University of Science and Technology of China, Shanghai, 201315, China}
\affiliation{CAS Center for Excellence and Synergetic Innovation Center in Quantum Information and Quantum Physics, University of Science and Technology of China, Hefei, Anhui 230026, China}
\affiliation{CAS-Alibaba Quantum Computing Laboratory, Shanghai, 201315, China}

\author{Jian-Wei Pan}
\affiliation{Shanghai Branch, National Laboratory for Physical Sciences at Microscale and Department of Modern Physics, University of Science and Technology of China, Shanghai, 201315, China}
\affiliation{CAS Center for Excellence and Synergetic Innovation Center in Quantum Information and Quantum Physics, University of Science and Technology of China, Hefei, Anhui 230026, China}
\affiliation{CAS-Alibaba Quantum Computing Laboratory, Shanghai, 201315, China}


\begin{abstract}
Recent experimental realizations of superfluid mixtures of Bose and Fermi quantum gases provide a unique platform for exploring diverse superfluid phenomena. We study dipole oscillations of a double superfluid in a cigar-shaped optical dipole trap, consisting of $^{41}$K and $^{6}$Li atoms with a large mass imbalance, where the oscillations of the bosonic and fermionic components are coupled via the Bose-Fermi interaction. In our high-precision measurements, the frequencies of both components are observed to be shifted from the single-species ones, and exhibit unusual features. The frequency shifts of the $^{41}$K component are upward (downward) in the radial (axial) direction, whereas the $^{6}$Li  component has down-shifted frequencies in both directions. Most strikingly, as the interaction strength is varied, the frequency shifts  display a resonant-like behavior in both directions, for both species, and around a similar location at the BCS side of fermionic superfluid. These rich phenomena challenge theoretical understanding of superfluids.
\end{abstract}

\maketitle

\date{\today}
The past two decades have witnessed vast experimental progress in generating and manipulating ultracold quantum gases, which have emerged as a powerful tool for simulating many body physics and particularly diverse superfluid and superconductivity phenomena~\cite{Ketterle2002Nobellecture,Bloch2008RMP,Bloch2012}. Examples include weakly interacting Bose-Einstein condensates (BEC)~\cite{Anderson198,PhysRevLett.75.3969}, the topological Berezinskii-Kosterlitz-Thouless phase transition~\cite{Cataliotti2001science,Hadzibabic2006Nature}, emergent relativistic phenomena near the superfluid-to-Mott-insulator quantum critical point~\cite{Endres2012Nature,Greif2016Science}, and the crossover between a BEC and a Bardeen-Cooper-Schrieffer (BCS) superfluid in fermionic systems~\cite{Regal2004PRL,Zwierlein2004PRL}. Very recently, experimental realizations of Bose-Fermi double superfluids have been reported for $^6$Li-$^7$Li, $^6$Li-$^{41}$K and $^6$Li-$^{174}$Yb mixtures~\cite{FerrierBarbut1035,XingCanYao2016,Roy2017PRL}. This is an important achievement, taking into account that, in the study of liquid helium, the strong inter-isotope interactions have prevented the long-sought goal of realizing  simultaneous superfluidity in a $^4$He-$^3$He mixture~\cite{Tuoriniemi2002}. Many fascinating behaviors can emerge in this novel quantum matter owing to Bose-Fermi interactions, including topological p-wave Cooper pairing~\cite{Rysti2012PRB}, Bose-Fermi dark solitons~\cite{Tylutki2016}, and polaronic atom-trimer continuity~\cite{Yusuke2015PRL}. 
Experimentally, although the measured critical velocity of a $^6$Li-$^7$Li mixture can be mostly accounted for by a generalized Landau criterion, it has an unexplained strong reduction on the BEC side~\cite{Delehaye2015PRL}. For a $^{6}$Li-$^{41}$K double superfluid, quantized vortices, a hallmark of superfluidity, are simultaneously generated in both species, and several unconventional interaction-induced properties are observed~\cite{XingCanYao2016}.

The investigation of collective excitations is well known to be an important method for gaining insights into the physical properties of trapped BECs and strongly interacting Fermi gases~\cite{Stringari1996PRL,Jin1996PRL,Bartenstein2004,Altmeyer2007,JinyiZhang2012PRL}. The simplest collective dynamics is the dipole oscillation for the center-of-mass motion of all atoms. Therefore, characterizing the dipole oscillation would be the first step to understand interaction-induced effects in Bose-Fermi superfluid mixtures. If all of the bosonic and fermionic atoms have the same mass and experience the same trapping frequency, the dipole oscillation frequency is exactly given by the non-interacting result, i.e., the intra- and inter-component interactions play no role in the dipole oscillation. This is known as the Kohn theorem~\cite{Kohn1961PR,Stringari1999RMP}. In a $^6$Li-$^7$Li superfluid mixture~\cite{FerrierBarbut1035,Delehaye2015PRL}, the slightly different masses of the two isotopes break the dynamical symmetry, and the dipole oscillation of the bosonic $^7$Li component has a down-shifted frequency and a beating behavior in the oscillation amplitudes. These results are well explained by a phenomenological two-oscillator model with effective mean-field interactions~\cite{FerrierBarbut1035}. Similar behaviors have been reported for a $^6$Li-$^{174}$Yb superfluid mixture~\cite{Roy2017PRL}, in which the dynamical symmetry is heavily broken by the large mass imbalance.

In this Letter, we report on the study of the coupled collective dipole oscillations of  a  $^{6}$Li-$^{41}$K superfluid mixture  in a cigar-shaped  optical dipole trap. 
We perform precise measurements  of oscillations in both the axial and radial directions, and observe three features that have not been reported in the previous experiments~\cite{FerrierBarbut1035,Delehaye2015PRL,Roy2017PRL}. First, both the bosonic and fermionic atoms have oscillation frequencies shifted from the single-species ones, although the frequency changes of the  $^6$Li component are small and only up to one and a half percent. Second, the frequency shifts of the $^{41}$K component are opposite: downward (upward) in the axial (radial) direction, while the $^6$Li component has down-shifted frequencies in both directions. Most strikingly, as the system is tuned across the BEC-BCS crossover, the axial and radial frequency shifts both display a resonant-like behavior on the BCS side around $1/k_{\rm F}a_{\rm f}=-0.2$, where $k_{\rm F}$ is the Fermi momentum and $a_{\rm f}$ is the scattering length of the $^6$Li atoms. A phenomenological analysis, which takes into account the mean-field interactions and their effects on the density profiles of both species, can only qualitatively describe a small part of the experimental data. 

\begin{figure}[htbp]
  \centering
  \includegraphics[width=\columnwidth]{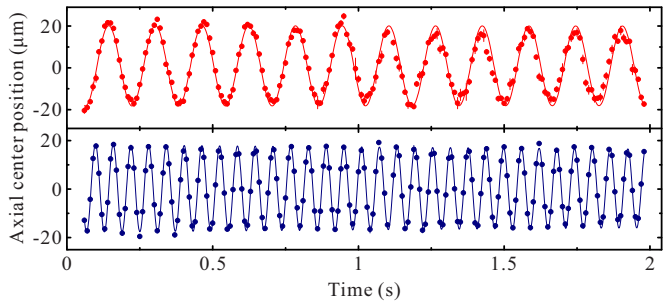}
  \caption{Axial dipole oscillations of the $^6$Li-$^{41}$K superfluid mixture at unitarity ($B=834$~G). 
  Symbols are the center-of-mass positions of  $^{41}$K (top) and $^6$Li (bottom) clouds, respectively. 
  Solid lines are the fitting curves according to an exponentially-damped sinusoid model.}\label{figure1}
\end{figure}

{\bf Experimental procedure.} The experimental procedure for preparing the Bose and Fermi superfluid mixture is similar to that of our previous works~\cite{Pan2017,XingCanYao2016,Chen2016PRA,Chen2016}.  After the laser cooling and magnetic transport phase, the mixture of cold $^6$Li and $^{41}$K atoms is confined in an optically plugged magnetic trap in a dodecagonal glass cell with good optical access and an ultrahigh vacuum environment, where radio frequency (rf) evaporation~\cite{Ketterle1996181} of $^{41}$K atoms is implemented and the $^6$Li atoms are cooled sympathetically. Then, we load the cold Bose-Fermi mixture into a cigar-shaped optical dipole trap (wavelength 1064~nm, $1/e^2$ radius 35~$\mu$m) and apply two 3~ms Landau-Zener sweeps to prepare both species at their lowest hyperfine states. Next, the magnetic field $B$ is ramped to 871~G, and a half-to-half spin mixture of the two lowest hyperfine states of $^6$Li is prepared using successive rf sweeps. After 0.5~s forced evaporation, the clouds are adiabatically transferred into the final cigar-shaped optical trap (wavelength 1064~nm, $1/e^2$ radius 62.5~$\mu$m) with large trap volume. Further evaporation is accomplished by exponentially lowering the trap depth to 512~nK ($^{41}$K) and 1.00~$\mu$K ($^6$Li) in 3~s. Finally, the Bose and Fermi superfluid mixture is achieved with approximately  $N_{\rm F}=1\times10^6$ $^6$Li atoms at 0.06(1) Fermi temperature and $N_{\rm B}=2.3\times10^5$ $^{41}$K atoms with a condensate fraction $\geq90\%$. In the gravity direction, the full overlap of the two species is achieved with a slightly off-centered distance of 10~$\mu$m at 871~G.

\begin{figure}[htbp]
  \centering
  \includegraphics[width=\columnwidth]{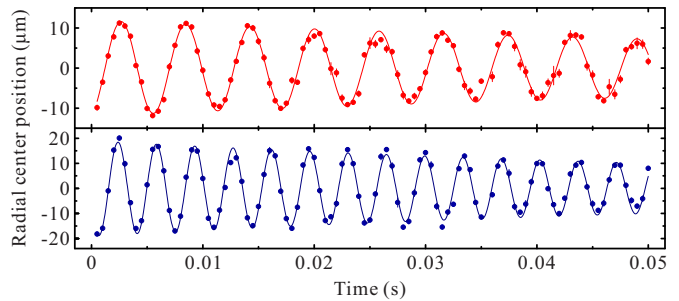}
  \caption{Radial dipole oscillations of the $^6$Li-$^{41}$K superfluid mixture at unitarity ($B=$834~G). 
  Symbols are the center-of-mass positions of  $^{41}$K (top) and $^6$Li (bottom) clouds, respectively.
  Solid lines are the fitting curves according to an exponentially-damped sinusoid model. }\label{figure3}
\end{figure}

At the end of evaporation, the magnetic field is ramped in 200~ms to the value for the oscillation experiment, and is held for another 500~ms to achieve fully thermal equilibrium of the two species. The dipole oscillation in the axial direction is induced by adiabatically shifting the center position of the superfluid mixture to a distance of approximately 18.5~$\mu$m in the weakly confining direction $z$,  and abruptly releasing the atom clouds in the trap. For the oscillation experiment in the radial direction,  the optical trap depth is adiabatically ramped up to 1.90~$\mu$K ($^{41}$K) and 1.80~$\mu$K ($^6$Li) in 200~ms after the Bose and Fermi superfluid mixture is achieved. The radial dipole mode is excited by adiabatically displacing the optical trap center by approximately 5~$\mu$m in 20~ms with an acoustical optical modulator and quickly shifting it back to the initial position in 100~$\mu$s. After a variable holding time, the optical trap potential is suddenly switched off, and the atom clouds are expanded for 2~ms in the residual magnetic curvature. A specially designed imaging setup in the gravity direction is employed to simultaneously probe the two species~\cite{XingCanYao2016}, and the centers of the atom clouds are simultaneously recorded. With a high numerical aperture objective, an imaging resolution of 2.2~$\mu$m (2.5~$\mu$m) at 671~nm (767~nm) is obtained.

{\bf Experimental result.} 
As an illustration, we focus on the unitary case ($B=834$~G). The recorded centers of the atom clouds are  shown in Figs.~\ref{figure1} and \ref{figure3}, respectively, for the axial and the radial directions, where the curves  can be nicely approximated by single-frequency harmonic oscillations. The damping rates are small, especially in the axial direction, where the oscillations of both species persist for more than 2~s without visible damping.

\begin{figure}[htbp]
  \centering
  \includegraphics[width=0.85\columnwidth]{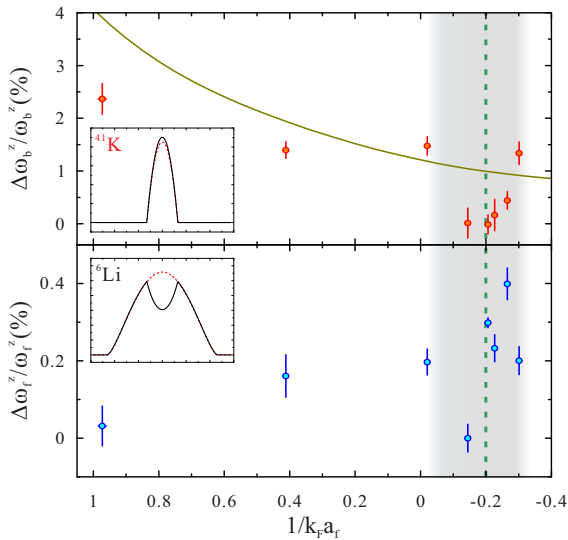}
  \caption{ Frequency shifts of axial dipole oscillations of the $^6$Li-$^{41}$K superfluid mixture in the BEC-BCS crossover. 
  Circles are measured frequency shifts of $^{41}$K (top, red) and $^6$Li (bottom, blue) atoms. Solid line is the theoretical predicted $^{41}$K frequency shift with mean-field model. 
  Green dashed line marks the position of $1/k_{\rm F} a_{\rm f}=-0.2$ and gray shadow shows the resonant-like region. 
  Error bars represent one standard deviation. 
  Insets are the numerically calculated density profiles of the $^{41}$K (top) and $^6$Li (bottom) cloud at 834~G,
  where the black solid lines are for superfluid mixture and the red dashed lines are for single-species superfluids. }\label{figure2}
\end{figure}

The experimental data are  fitted by exponentially damped sinusoidal models. For the axial oscillations, the fits yield frequencies  $\widetilde{\omega}_{\rm f}^z=  2\pi  \times 16.420(5)$~Hz for the fermionic $^6$Li component and $\widetilde{\omega}_{\rm b}^z=2\pi\times 6.202(7)$~Hz for the bosonic $^{41}$K atom cloud, with
a relative precision of up to $0.03\%$ for $^6$Li  and $0.11\%$ for $^{41}$K, respectively. The damping constants $\widetilde{\omega}_i^z\tau_i^z$ ($i={\rm b,f}$) are larger than $10^3$ for both species. In the radial direction,  the frequencies are $\widetilde{\omega}_{\rm f}^{r}=2\pi\times 291.1(4)$~Hz ($^6$Li) and $\widetilde{\omega}_{\rm b}^{r}=2\pi \times174.3(4)$~Hz ($^{41}$K), and the damping constants are $\widetilde{\omega}_{\rm f}^r\tau_{\rm f}^r\simeq142$ 
and $\widetilde{\omega}_{\rm b}^r\tau_{\rm b}^r\simeq110$, respectively. The fitting results are shown as the solid lines in Figs.~\ref{figure1} and \ref{figure3}. 
Nearly all of the experimental data points lie on top of the fitting curves or have small deviations. This demonstrates the stability of our experimental setup as well as the fitting quality.

Comparison studies of the dipole oscillations are carried out for the single-species superfluids, consisting only of $^6$Li or $^{41}$K atoms. We obtain oscillation frequencies $\omega_{\rm f}^z= 2\pi\times 16.453(2)$~Hz ($^6$Li) and $\omega_{\rm b}^z=2\pi\times 6.295(8)$~Hz ($^{41}$K) in the axial direction, and 
$\omega_{\rm f}^r=2\pi\times 295.4 (3) $~Hz ($^6$Li) and $\omega_{\rm b}^r =2\pi\times 170.7(5) $~Hz ($^{41}$K) in the radial direction. The relative frequency shift is then computed as $\Delta \omega /\omega \equiv (\omega-\widetilde{\omega})/\omega$, with positive (negative) values for downward (upward) shifts. We obtain $\Delta\omega_{\rm b}^z/\omega_{\rm b}^z =+1.5(2)\%$  ($^{41}$K) and $\Delta\omega_{\rm f}^z/\omega_{\rm f}^z=+0.20(3)\%$ ($^6$Li) for the axial oscillations, and $\Delta\omega_{\rm b}^r/\omega_{\rm b}^r =-2.1(4)\%$ and $\Delta\omega_{\rm f}^r/\omega_{\rm f}^r =+1.5(2)\%$  for the radial oscillations.

\begin{figure}[htbp]
  \centering
  \includegraphics[width=0.85\columnwidth]{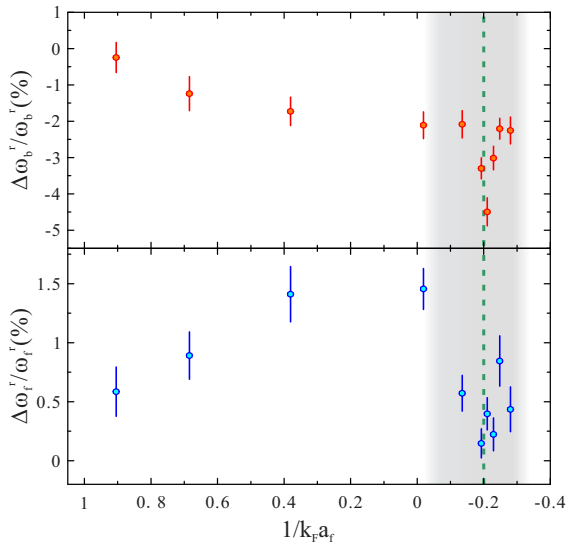}
  \caption{Frequency shifts of radial dipole oscillations of the $^6$Li-$^{41}$K superfluid mixture in the BEC-BCS crossover. Circles are measured frequency shifts of $^{41}$K (top, red) and $^6$Li (bottom, blue) atoms. Green dashed line marks the position of $1/k_{\rm F} a_{\rm f}=-0.2$ and gray shadow shows the resonant-like region. 
  Error bars represent one standard deviation.}\label{figure4}
\end{figure}

By applying the same procedures for the experiment and data analyses, we study the coupled dipole oscillations across the BEC-BCS crossover in the range of [767~G, 862~G] for the magnetic field. The damping rates are small in the entire parameter regime: the damping constants are bigger than $10^3$ in the axial direction and are  between 100 and 200 for the radial oscillations. The results for the frequency shifts are shown in Figs.~\ref{figure2} and ~\ref{figure4}. They display a few interesting features that have not been reported in the previous experiments~\cite{FerrierBarbut1035,Delehaye2015PRL,Roy2017PRL}.

First, it is confirmed that the Bose-Fermi interaction can lead to frequency changes in both the bosonic and fermionic superfluid components. Note that, since  the cloud size of the bosonic atoms is  smaller than that of the fermionic atoms, it is more difficult to induce a frequency shift in the latter. Indeed, no frequency change was observed in the dipole oscillations of the fermionic $^6$Li cloud in $^6$Li-$^7$Li~\cite{FerrierBarbut1035,Delehaye2015PRL}. In our case, the $^6$Li atom cloud has a smaller frequency shift  than the $^{41}$K component (Figs.~\ref{figure2} and~\ref{figure4}). Nevertheless, the existence of frequency shifts is well supported by the fact that the experimental data are far beyond the statistical uncertainties. We attribute the frequency shifts in the fermionic component to the large mass imbalance of the  $^{41}$K and $^6$Li atoms, which can lead to pronounced interaction effects. For instance, it is observed that the repelling from the $^{41}$K superfluid can induce a density-profile depression in the center of the $^6$Li component~\cite{XingCanYao2016}.  

Second,  the dipole oscillation of the bosonic $^{41}$K component has an up-shifted frequency in the radial direction and a down-shifted one in the axial direction. As shown in Ref.~\cite{FerrierBarbut1035}, the frequency downshift can be straightforwardly explained using a phenomenological model with mean-field interactions.  
However, it is not clear what leads to the frequency upshifts. A possible explanation is scattering between normal component in the superfluid mixture, such a dissipation-induced coupling might drive the two species to oscillate with intermediate frequencies. The radial dipole oscillations have a frequency that is higher than the axial one by about 20 times, and are expected to suffer from more severe thermal effects. 

Third, as the interaction strength is varied, a resonant-like behavior emerges in both the axial and radial oscillations, for both the Bose and Fermi species, 
and at a similar location of approximately $1/k_{\rm F}a_{\rm f}=-0.2$ (BCS side) within a window of a size about $0.2$.  
The radial and axial frequency shifts of the  $^{41}$K component  exhibit a pronounced dip near  $1/k_{\rm F}a_{\rm f}=-0.2$, where the axial downshift drops to nearly zero and the radial upshift reaches a maximal value of about $4.5\%$. The frequency shifts of the fermionic $^6$Li superfluid also display a non-smooth behavior around $1/k_{\rm F}a_{\rm f}=-0.2$. We mention that the damping rates remain approximately constant in the BEC-BCS crossover for both the axial and radial oscillations.

{\bf Phenomenological  analysis.} 
In the previous works~\cite{FerrierBarbut1035,Delehaye2015PRL,Roy2017PRL}, the frequency downshifts in the bosonic component are quantitatively described using a simple mean-field model. The bosonic component is treated as a mesoscopic impurity immersed in a Fermi superfluid, and the effective potential  seen by the Bosons is the sum of the trapping potential and the mean-field Bose-Fermi interaction. With the local density approximation, this leads to a harmonic potential with a reduced frequency. However, for the $^6$Li-$^{41}$K superfluid mixture, this model can be improved since the density distribution of the $^6$Li cloud is significantly altered by the Bose-Fermi interaction~\cite{XingCanYao2016}. 

Here, we adopt an improved mean-field model~\cite{Tomoki2014PRA,WenWen2017JPB}, which takes into account the interaction-induced alternation of the density distributions for both species. On the basis of the coupled hydrodynamic description, the density distribution of the superfluid mixture in the overlapping region can be expressed as~{\cite{WenWen2017JPB}},
\begin{equation}\label{equ2}
\begin{aligned}
n_{\rm b}(\mathbf{r})=& (1/g_{\rm b}) [\mu_{\rm b}-V_{\rm b}(\mathbf{r})-2g_{\rm bf} n_{\rm f}(\mathbf{r})] \; ,\\
n_{\rm f}(\mathbf{r}) =& n_{\rm f}^0 \{ (1/g_{\rm b} \mu_{\rm f}^0) [ g_{\rm b} \mu_{\rm f}-g_{\rm bf} \mu_{\rm b}
-g_{\rm b} V_{\rm f}(\mathbf{r}) \\
&+g_{\rm bf} V_{\rm b}(\mathbf{r})+2 g_{\rm bf}^2 n_{\rm f}(\mathbf{r})] \}^{1/\gamma}
\end{aligned}
\end{equation}
where $n_{\rm f}^0$ and $\mu_{\rm f}^0$ are the reference particle number density and the chemical potential of the $^6$Li atoms, respectively,
$g_{\rm bf}=2\pi \hbar^2 a_{\rm bf}/m_{\rm bf}$ and $g_{\rm b}=4\pi\hbar^2 a_{\rm b}/m_{\rm b}$ are the Bose-Fermi  and the Bose-Bose interaction coupling constants, respectively, and $m_{\rm bf}=m_{\rm b} m_{\rm f}/(m_{\rm b}+m_{\rm f})$ is the $^6$Li-$^{41}$K reduced mass.
The scattering lengths are $a_{\rm bf}=60.2a_0$ and $a_{\rm b}=60.5a_0$ ($a_0$ is the Bohr radius)~\cite{XingCanYao2016}. Symbol $\gamma$ is the polytropic index of the equation of state of the strongly interacting Fermi gas~\cite{Menotti2002PRL,Navon2010Nature,wu2017oscillatory}. At unitarity ($B=834$~G),  the calculated static density distributions are shown in the insets of Fig.~\ref{figure2}, clearly displaying a density depression in the $^6$Li atom cloud. Note that the maximum relative velocity  is far below the critical velocity, e.g., $v_{\rm max}^z$ $\simeq$  2.5~mm/s $<< v_{c, \rm{f}}^z $$\simeq$ 17~mm/s at the unitary limit~\cite{PhysRevLett.114.095301,Delehaye2015PRL}, and that the depression in the $^6$Li density should adiabatically follow the movement of the $^{41}$K cloud. The effective potential seen by the $^{41}$K atoms is then expressed as
\begin{equation}\label{equ3}
  \widetilde{V}_{{\rm b}}(\mathbf{r}_1)=\frac{\int (V_{\rm b}(\mathbf{r}+\mathbf{r}_1)+2g_{\rm bf}n_{\rm f}(\mathbf{r},\mathbf{r}_1))\times n_{\rm b}(\mathbf{r})d\mathbf{r}}{\int n_{\rm b}(\mathbf{r})d\mathbf{r}}
\end{equation}
where $\mathbf{r}_1$ is the center-of-mass position of the $^{41}$K BEC,  and $n_{\rm f}(\mathbf{r},\mathbf{r}_1)$ is the $\mathbf{r}_1$-dependent density distribution of the $^6$Li superfluid. The oscillation frequency of the $^{41}$K component  is given by $\widetilde{\omega}_{\rm b}^z=\sqrt{(\frac{\partial^2\widetilde{V}_{{\rm b}}(\mathbf{r}_1)}{\partial z^2}|_{\mathbf{r}_1=0})/m_{\rm b}}$. The resulting frequency downshifts, shown as the solid line in Fig.~\ref{figure2}, exhibit similar trends and decrease from the BEC to BCS sides of the crossover. The deviations on the BEC side might be caused by the deformation of the $^{41}$K density profile during oscillations, where the $^6$Li density is significantly enhanced compared to the BCS regime.

In conclusion, our experiments demonstrate that the coupled dipole oscillations of the  $^6$Li-$^{41}$K superfluid mixture, the simplest collective mode, show 
a variety of rich behaviors, particularly a pronounced dependence on the Bose-Fermi interacting strength. Despite that a phenomenological mean-field description can qualitatively explain the downward frequency shifts in the axial oscillations of the bosonic component, we cannot account for in the mean-field framework the upshifts of $^{41}$K  in the radial direction or the frequency shifts of the $^6$Li component. The most striking feature is the resonant-like frequency change of the dipole oscillations in the regime of a strongly attractive Fermi gas with $1/k_{\rm F}a_{\rm f}=-0.2$. An immediate question is: do the similar locations for both the axial and radial directions imply some unknown universal mechanism? More careful experimental studies as well as theoretical investigations are needed, e.g., to investigate the temperature dependence effect. Besides providing valuable information on the physical properties of the system, our experimental results of the coupled dipole oscillations offer a promising prospect that a Bose-Fermi superfluid mixture with a large mass imbalance may exhibit rich quantum phases and phase transitions.

\begin{acknowledgements}
We are indebted to valuable discussions with C. Salomon and H. Zhai. This work has been supported by the NSFC of China, the CAS, and the National Fundamental 
Research Program (under Grant Nos. 2013CB922001 and 2016YFA0301600). 

Y.-P. Wu, X.-C. Yao and X.-P. Liu contributed equally to this work.
\end{acknowledgements}

%

%

\end{document}